\documentclass{article}

\begin{document}

\bigskip

\bigskip

\bigskip

\begin{center}
{\Large\bf Evaluation of neutrino masses from 
$m_{\beta\beta}$ values}

\bigskip

\bigskip

{\large V.V. Khruschov}

\smallskip

{\small\it Russian Research Centre "Kurchatov Institute",
Kurchatov Sq., 123182 \\ Moscow, Russia (e-mail:khru@imp.kiae.ru)}
\end{center}

\vspace{-7mm}

\begin{abstract}

A neutrino mass matrix is considered under conditions of a CP invariance and 
a small reactor mixing $\theta_{13}$ angle. 
Absolute mass values for three neutrinos are evaluated
for normal and inverted hierarchy spectra  on the ground of data
for oscillation mixing neutrino parameters and   an
effective neutrino mass $m_{\beta\beta}$ related to a probability of 
neutrinoless two beta decay.
\end{abstract}








\bigskip


{\bf 1. Introduction }

\smallskip

Discovering of neutrino oscillations, which have been predicted
in the Refs. \cite{1,2}, points to a nonconservation of lepton quantum
numbers $L_e$, $L_{\mu}$, $L_{\tau}$ and nonequal each other neutrino
masses. Now further objectives for neutrino physics investigations are a
clarification of Dirac or Majorana nature for neutrinos, a refinement
of neutrino mixing parameters and a determination of neutrino mass absolute
values \cite{3,4}.

Different theoretical approaches may be used to solve these problems, for
instance, Grand Unification Theories (GUTs) or various phenomenological
models (PMs). The characteristical feature for many neutrino physics articles 
is the use of continuous or discrete symmmetries beyond the known 
symmmetries of the Standard Model (SM). For instance, 
absolute mass values of three neutrino were  obtained
in the framework of the  three-flavor model for Majorana neutrinos with
the Pauli symmetry ($0,054\, eV <m_{1,2} < 0, 391\, eV$,
 $ 0,058\, eV <m_{3} < 0,406\, eV$) \cite{5}.
In the PM context interesting results were also obtained:
the phenomenological scheme of three-bimaximal mixing of neutrinos, where
the explicit form (see Sec.4) of neutrino mixing matrix was given in 
Ref.\cite{6};
the relations between quark and neutrino mixing angles \cite{7},
which, e.g., were used in Ref. \cite{8} for precise neutrino mixing 
angles evaluation; constraints on the neutrino mass matrix structure
and dependences between the minimal neutrino mass and the
effective neutrino mass related to a probability of neutrinoless
two beta decay $m_{\beta\beta}$, these dependences were presented
in the graphical form in Refs.\cite{9,10}.
There are many papers \cite{11,12,13,14} about  a 
 neutrino mass matrix structure, in particular about a number
and  locations of zeroth matrix elements.

The present paper is devoted to an examination of a neutrino mass 
matrix structure and an elucidation a role of some assumptions, 
which make possible to find absolute neutrino mass values 
on  the basis of the minimally extended SM with three types of massive
neutrinos \cite{3,4}. The set of existing experimental data 
concerning neutrino oscilations \cite{15} has a basic
role in obtaining of neutrino mass values. In Sec.2 a general 
neutrino mass matrix structure for  Majorana neutrinos
in the flavor representation is presented together with its
link to the neutrino mixing matrix under the CP invariance condition.
In Sec. 3 additional simple assumptions  are considered,
which facilitate  neutrino mass matrix structure and allow 
to evaluate absolute neutrino masses at specified $m_{\beta\beta}$
values. In doing so  neutrino masses became dependent on
mixing angles. It should be noted that the main assumption, that used
throughout the present paper, is the condition of 
the small reactor mixing  $\theta_{13}^{\nu}$ angle, that is in 
accordance with the experimental data (e.g., see \cite{15,22}).
 In Sec.4  neutrino mass evaluations are
 performed at prescribed $m_{\beta\beta}$ values from an acceptable
range and experimentally determined oscilation mixing parameters of
neutrinos. As a check on performed calculations
the limiting case is treated when $m_{\beta\beta}$ equals to zero and 
the three-bimaximal neutrino
mixing takes place. In the last section the results obtained 
are discussed and their possible generalization for nonzero 
$\theta_{13}^{\nu}$ values are considered.

\medskip

{\bf 2. Neutrino mixing matrix and neutrino mass matrix}

\smallskip

As it is known, left components of fields of flavor $\nu_{\alpha}, 
\alpha = e, \mu, \tau$ and massive $\nu_{i}, i = 1, 2, 3$ neutrinos are 
connected with the Pontecorvo-Maki-Nakagava-Sakata matrix:
\begin{equation}\label{}
    \nu_{\alpha L}= \sum_{i=1}^{3}U_{PMNS,\alpha i} \nu_{i L}
\end{equation}

In general case when neutrinos are Majorana particles the 
$U_{PMNS}$ matrix may be specified in the form: $U_{PMNS}^D\cdot P^M$,
where matrix $U_{PMNS}^D$ coincides with the mixing matrix
of Dirac neutrinos, while $P^M$ is the diagonal matrix containing
additional phases due to CP noninvariance in processes involving
Majorana neutrinos. Let us write the $U_{PMNS}$ matrix in the
standard parameterization apart from  new notations for mixing angles
and phases, which differ from  angles
and phases notations in Cabibbo-Kobayashi-Maskava matrix.
The Dirac phase  $\delta$ is denoted as $\epsilon$, angles
$\theta_{ij}^{\nu}$ are denoted as $\eta_{ij}$, so
$c_{ij}\equiv\cos\eta_{ij}$ $s_{ij}\equiv\sin\eta_{ij}$, while
 $\phi_1$, $\phi_2$ are the Majorana phases.
$$
   U_{PMNS}=\left(
  \begin{array}{ccc}
    c_{12}c_{13} & s_{12}c_{13} & s_{13}e^{-i\epsilon} \\
    -s_{12}c_{23}-c_{12}s_{23}s_{13}e^{i\epsilon} & c_{12}c_{23}-s_{12}s_{23}s_{13}e^{i\epsilon} & c_{13}s_{23} \\
    s_{12}s_{23}-c_{12}c_{23}s_{13}e^{i\epsilon} &-c_{12}s_{23}-s_{12}c_{23}s_{
    13}e^{i\epsilon}  & c_{13}c_{23} \\
  \end{array}
\right)\cdot $$
\begin{equation}
\left(
               \begin{array}{ccc}
                 e^{-i\phi_1} & 0 & 0 \\
                 0 & e^{-i\phi_2} & 0 \\
                 0 & 0& 1 \\
               \end{array}
             \right)
\end{equation}

It is common knowledge that at present experimental data concerning
 phases $\epsilon$, $\phi_1$ and $\phi_2$ are absent. Moreover,
these phases will not be observable for a long period of time due to 
the small $\eta_{13}$ value and difficulties in realization 
of experiments for the $\phi_1$ and $\phi_2$ determination.
Thus we consider processes when the CP invariance holds and the reactor
mixing angle $\eta_{13}$ is negligible. In this case the
neutrino mass matrix can be presented in the flavor basis in 
the form:
\begin{equation}\label{}
    M^{\nu}_f = U_{PMNS} M^{\nu}_m U_{PMNS}^T,
\end{equation}
\noindent where $M^{\nu}_m=diag\{m_1, m_2, m_3\}$.  $m_i, i=1,2,3,$ can have
minus and plus signs, and the $U_{PMNS}$ matrix depends only on two
mixing angles, namely $\eta_{12}$ and $\eta_{23}$. Note that 
the minus and plus $m_i$ signs are associated with relative CP parities of
neutrinos (e.g. see \cite{3,4}). Let us write the explicit form of
the neutrino mass matrix taking into account the constraints,
which have been made.
$$
 \hspace*{-0.3cm}  M^{\nu}_f=\left(
  \begin{array}{cc}
    m_1c_{12}^2+m_2s_{12}^2 & -m_1c_{12}s_{12}c_{23}+m_2c_{12}s_{12}c_{23}  \\
    -m_1s_{12}c_{23}c_{12}+m_2c_{12}c_{23}s_{12} & m_1s_{12}^2c_{23}^2
+m_2c_{12}^2c_{23}^2+m_3s_{23}^2 \\
   m_1c_{12}s_{12}s_{23}-m_2c_{12}s_{12}s_{23} &-m_1s_{12}^2c_{23}s_{23}-
m_2c_{12}^2c_{23}s_{23}+m_3c_{23}s_{23} \\
  \end{array}
\right. $$
\begin{equation}
\left.
       \begin{array}{c}
       \hspace*{4cm}  m_1c_{12}s_{12}s_{23}-m_2c_{12}s_{12}s_{23}  \\
\hspace*{4cm}-m_1s_{12}^2c_{23}s_{23}-m_2c_{12}^2s_{23}c_{23}+m_3s_{23}c_{23}\\
  \hspace*{4cm}  m_3c_{23}^2+m_1s_{12}^2s_{23}^2+m_2c_{12}^2s_{23}^2  \\
               \end{array}
             \right)
\end{equation}

It is well known that various assumptions are considered concerning matrix 
elements values in order to specify a particular structure of the 
$M^{\nu}_f$ matrix  \cite{11,12}. For instance, conditions of zero values
for a few matrix elements or the spur and the deternimant
are used. However in the present paper we assume that 
the spur and the deternimant of the $M^{\nu}_f$ matrix are not equal to zero. 
Besides we use the condition, that a value of the first
diagonal $M^{\nu}_f$ matrix element (or some restriction on this value)
can be determinated in principle in experiments for a neutrinoless 
two beta decay search.

\medskip

{\bf 3. Additional restrictions on mixing
angles and neutrino mass spectra.
}

\smallskip

As indicated above, in the neutrino PM framework
additional assumptions on the mass matrix $M^{\nu}_f$ 
or its matrix elements are used in order to obtain some
results to be tested in experiments. For example, in the paper \cite{11}
the case when the modulus of the diagonal matrix element $M^{\nu}_{ee}$
equals zero is investigated in detail. In Refs. \cite{13,14} cases have
been considered when the spur $SpM^{\nu}$ or the determinant 
$DetM^{\nu}$ 
\begin{equation}\label{}
    SpM^{\nu}= m_1+m_2+m_3,\quad DetM^{\nu}= m_1m_2m_3.
\end{equation}
\noindent are equal to zero values.
In the present paper we put that the spur and the determinat for the
$M^{\nu}_f$  matrix are not equal to zero.
\begin{equation}\label{}
    SpM^{\nu}\neq 0,\quad DetM^{\nu} \neq 0.
\end{equation}

It is convenient to use the following classification for 
possible neutrino mass spectra: the case for  $m_1 < m_2 \ll m_3$
will be called the neutrino mass spectrum with a normal hierarchy (NH),
the case for  $m_3\ll m_1<m_2$ 
will be called the neutrino mass spectrum with an inverted hierarchy (IH).

It is known a value of the first diagonal matrix element
$M^{\nu}_{ee}$ is connected with a probability of a  neutrinoless
two beta decay of nucleus: $(A,Z) \to (A, Z+2) + 2e$. Actually the half-life
of  $0\nu 2\beta$ decay  $T_{1/2}^{0\nu 2\beta}$ is in an inverse proportion
to the square of a $M^{\nu}_{ee}$ modulus. For this reason the 
$M^{\nu}_{ee}$ modulus
is usually denoted as $m_{\beta\beta}$ \cite{Avi}. A search for a neutrinoless
two beta decay is intended to reveal Dirac or Majorana nature of neutrinos.
 Moreover, a discovery of 
this decay makes possible to deternime an absolute scale of neutrino masses.
The conditions for picking of isotopes for a reliable detection of a 
signal of a neutrinoless two beta decay above the total background 
were considered in Ref.\cite{Zeld}.

If one use the parameterization (2) provided CP invariance holds, then
the matrix element $M^{\nu}_{ee}$ modulus  can be written in the form:
\begin{equation}
m_{\beta\beta}\equiv |M^{\nu}_{ee}| =|m_1c_{12}^2c_{13}^2+m_2s_{12}^2c_{13}^2+
m_3s_{13}^2|,
\label{mbebe}
\end{equation}
however as $\eta_{13}\approx 0$, hence 
 $ m_{\beta\beta} \approx |m_1c_{12}^2+m_2s_{12}^2|$.

Results of the experiments performed (Heidelberg-Moscow, IGEX)
provide the upper limit for admissible $m_{\beta\beta}$ values (90\% C.L.): 
$m_{\beta\beta}<0,35\div1,05 eV$, $m_{\beta\beta}<0,33\div1,35 eV$,
respectively \cite{16,17,18}. On the other hand
there are data for oscillation neutrino mixing parameters, which
were obtained in many experiments \cite{15}, such as the 
differencies of squares of neutrino masses $\Delta m_{ij}^2$ 
and mixing angles $\theta_{ij}$:
$$
 \Delta m_{21}^2 = (7,7\div8,3)\times 10^{-5}eV^2, 
$$
$$
|\Delta m_{32}^2|= (1,9\div3,0)\times 10^{-3}eV^2,
$$
$$
\sin^2(2\theta_{12}) = 0.86(+0.03/-0.04), 
$$
\begin{equation}
 \sin^2(2\theta_{23})>0.92,\quad \sin^2(2\theta_{13})< 0.19
\label{exp}
\end{equation}

\medskip

{\bf 4. Evaluation of absolute values for neutrino masses}

\smallskip

It is possible to evaluate absolute values for neutrino masses using at
 $\eta_{13} = 0$ the relation (\ref{mbebe}), experimental restrictions
 (\ref{exp}) as well as fixed ranges for  $m_{\beta\beta}$ values.
In the Table 1 neutrino mass values are presented
in the five ranges for  $0\leq m_{\beta\beta} \leq 0,9 eV$,
where $m^1_3$, $m^2_3$ denote the $m_3$ masses for the IH spectrum and
the NH spectrum, correspondingly. Performing $m_i$, $i=1,2,3$, masses 
evaluations the ranges
of acceptable $m_{\beta\beta}$ values have been setted, in so doing the 
possibility of different $m_i$ signs have been taken into account.

As a check on performed calculations one can use 
the limiting case  when the matrix element
$M_{ee}^{\nu}$ equals to zero and the three-bimaximal neutrino
mixing takes place \cite{1,2}. As is known, experimental data  for 
mixing angles are approximated well with the help of the three-bimaximal
matrix:

\begin{equation}\label{TBM}
    U^{TBM}_{PMNS}=
    \left(
      \begin{array}{ccc}
        2/\surd 6 & 1/\surd 3 & 0 \\
        -1/\surd 6 & 1/\surd 3 &1/\surd 2 \\
        1/\surd 6 & -1/\surd 3 & 1/\surd 2 \\
      \end{array}
    \right)
\end{equation}

\noindent If one use the condition $m_{\beta\beta}\approx 0$ as an admissible
approximation together with the $U^{TBM}_{PMNS}$ matrix (\ref{TBM}),
then the following neutrino mass values can be obtained:
\begin{equation}\label{}
    \begin{array}{r}
        m_{1}\approx (5,1\div5,3)\times 10^{-3}eV,\\
       m_{2}\approx (10,1\div10,5)\times 10^{-3}eV, \\
       m_{3}\approx (44,7\div55,8)\times 10^{-3}eV
    \end{array}
\label{tbmm}
\end{equation}

\newpage

{\it Table 1. Neutrino masses evaluated against permissible
 $m_{\beta\beta}$ values}


\begin{center}
\begin{tabular}{|c|c|c|c|c|}
  \hline
  {} & {} & {} & {} & {} \\
  $m_{\beta\beta}$, eV & $m_1$, eV & $m_2$, eV & $m^1_3$, eV & $m^2_3$, eV \\
  {} & {} & {} & {} & {} \\\hline
  {} & {} & {} & {} & {} \\
  $(5\pm4)\cdot10^{-1}$ & $0.1\div0.9$ & $0.1004\div0.9001$ & $0.084\div0.899$& $0.11\div0.902$ \\
  {} & {} & {} & {} & {} \\\hline
  {} & {} & {} & {} & {} \\
   $(5\pm4)\cdot10^{-2}$& $0.009\div0.497$ & $0.0124\div0.4971$ & $0.\div0.495 $& $0.0453\div0.5$\\
  {} & {} & {} & {} & {} \\\hline
  {} & {} & {} & {} & {} \\
  $(5\pm4)\cdot10^{-3}$ & $0.\div0.029$ & $0.009\div0.03 $& $ - $& $0.044\div0.062$\\
  {} & {} & {} & {} & {} \\\hline
  {} & {} & {} & {} & {} \\
  $(5\pm4)\cdot10^{-4}$ & $0.004\div0.007$ & $0.0096\div0.012$ & $ - $ & $0.045\div0.056$\\
  {} & {} & {} & {} & {} \\\hline
  {} & {} & {} & {} & {} \\
  $(5\pm4)\cdot10^{-5}$& $0.004\div0.005 $& $0.0096\div0.011 $& $ - $ & $0.045\div0.056$\\
  {} & {} & {} & {} & {} \\\hline
  {} & {} & {} & {} & {} \\
  $\approx 0$ & $0.004\div0.005$ & $0.0096\div0.011 $& $ - $   & $0.045\div0.056$\\
  {} & {} & {} & {} & {} \\  \hline
\end{tabular}
\end{center}
\smallskip

It should be noted, that in the present paper the effective 
neutrino mass $m_{\beta\beta}$ for a neutrinoless two beta decay
is used as the independent parameter in order to find  $m_i$, $i=1,2,3$
absolute values, while an inverse dependence is treated in Refs.
\cite{9,10}, that was displayed in the graphical  form.
The correspondence between the results of the Refs. \cite{9,10} and
the present paper exists within the accuracy dependent on the accuracy
of graphical representation and employed experimental data. Furthermore
the admissible range for  $m_{\beta\beta}$ was extended here and
numerical results are more convenient in some cases  and give more
precise  $m_i$, $i=1,2,3$ values.

\medskip

{\bf 5. Conclusions and discussion }

\smallskip
 
As may be seen by comparison the mass values (\ref{tbmm}) 
and the mass values presented in the Tabl.1 the  evaluations of
 absolute neutrino masses are rather precise at small 
$m_{\beta\beta}$ values in despite of taking into account
uncertainties for experimental data. The values obtained are in accordance 
with the results of the Refs.\cite{9,10}
within the accuracy dependent on the  graphics presented and  data
used, along with the results of the Ref.\cite{18}, where conditions are
considered which cause  $m_{\beta\beta}$ values
larger than  $10^{-3}$ eV. This value is critical, it will be seen
from the reasoning below. As apparent from the Tabl.1 
the IH spectrum can not be realized at
$m_{\beta\beta}<0.01 eV$, this fact is previously found in Ref.\cite{9}.

One can see, when $m_{\beta\beta}<10^{-3} eV$, the neutrino mass values
practically are insensitive to $m_{\beta\beta}$ values. The values are
$m_1 = 0.0045 eV$, $m_2 = 0.0103 eV$, $m_3 = 0.0505 eV$. The same 
values of  $m_i$, $i=1,2,3$, have been obtained in  Ref.\cite{fri}.
Due to this fact there is a possibility to take into account small
but nonzero values of $\eta_{13}$ for the NH mass spectrum. In this 
case  the small values of $ m_3s_{13}^2$, that is less than 
$10^{-3} eV$, cannot change $m_i$ values either. So we obtain
the following condition $ s_{13}^2<2\cdot10^{-2}$, under this condition
and the previous condition: $m_{\beta\beta}<10^{-3} eV$, the neutrino
masses are equal to the invariable values.

The obtained estimations of neutrino masses at prescribed $m_{\beta\beta}$
values can be used for planning of experiments for a neutrinoless
two beta decay search and for  interpretation of results obtained. They
can also be used for  interpretation of any experimental results depended on 
absolute values of neutrino masses. For instance, the experiments in 
Troizk and Mainz for the measurement of a electron spectrum form in the 
 $\beta$ decay of tritium give the limit for a antineutrino mass
$m(\bar\nu_e)<2.2 eV$ (95\% C.L.) \cite{19,20}, that is expected will be
 $0.2 eV$ in the planned  KATRIN experiment\cite{21}. 
A statistical analysis of future neutrino mass experiments including
neutrinoless double beta decay have been performed in Ref.\cite{man} in order
to check possibility of a reconstruction of a type of a neutrino mass spectrum.

Results of astrophysical
and cosmological experiments lead to the following restriction
on the sum of neutrino masses  $\Sigma<0.19 эВ$ \cite{22}. It is 
significant that the domain of very small 
$\eta_{13}$  and $m_{\beta\beta}$ values is practically accessible only
for a theoretical examination. In the future we suppose to generalize
 the considered method for evaluation of neutrino masses, for instance,
 by taking into account effects of a CP noninvariance and  
sufficiently large  $\eta_{13}$ values, it makes possible
 to compare it
 with  other methods for evaluation of absolute values of 
neutrino masses, in particular, with the  three-flavor model 
for Majorana neutrinos with Pauli symmetry \cite{5}. Note that
allowing for a CP noninvariance can change the results obtained
for Majorana neutrinos even with small $\eta_{13}$ values.

Author is grateful to Yu.V. Gaponov and S.V. Semenov
for stimulating and useful discussions. The work was supported with
the grant of RRC "Kurchatov Institute" on fundamental research
\# 33 in 2008 year.


\end{document}